\documentclass[%
reprint,
superscriptaddress,
 amsmath,amssymb,
 aps,
pra,
]{revtex4-2}

\usepackage{graphicx}
\usepackage{dcolumn}
\usepackage{bm}
\usepackage{siunitx}

\begin{document}

\preprint{APS/123-QED}

\title{Tuning Wave-Particle Duality of Quantum Light by Generalized Photon Subtraction}

\author{Kan Takase}
\email{kan.takase@optqc.com}
\affiliation{Department of Applied Physics, School of Engineering, The University of Tokyo, Tokyo 113-8656, Japan}
\affiliation{Optical Quantum Computing Research Team, RIKEN Center for Quantum Computing, Wako, Saitama 351-0198, Japan}
\affiliation{OptQC Corp., 1-21-7 Nishi-Ikebukuro, Toshima, Tokyo, Japan}

\author{Mamoru Endo}
\email{endo@ap.t.u-tokyo.ac.jp}
\affiliation{Department of Applied Physics, School of Engineering, The University of Tokyo, Tokyo 113-8656, Japan}
\affiliation{Optical Quantum Computing Research Team, RIKEN Center for Quantum Computing, Wako, Saitama 351-0198, Japan}

\author{Fumiya Hanamura}
\affiliation{Department of Applied Physics, School of Engineering, The University of Tokyo, Tokyo 113-8656, Japan}

\author{Kazuki Hirota}
\affiliation{Department of Applied Physics, School of Engineering, The University of Tokyo, Tokyo 113-8656, Japan}

\author{Masahiro Yabuno}
\affiliation{Advanced ICT Research Institute, National Institute of Information and Communications Technology, 588-2 Iwaoka, Nishi-ku, Kobe, Hyogo 651-2492, Japan}

\author{Hirotaka Terai}
\affiliation{Advanced ICT Research Institute, National Institute of Information and Communications Technology, 588-2 Iwaoka, Nishi-ku, Kobe, Hyogo 651-2492, Japan}

\author{Shigehito Miki}
\affiliation{Advanced ICT Research Institute, National Institute of Information and Communications Technology, 588-2 Iwaoka, Nishi-ku, Kobe, Hyogo 651-2492, Japan}

\author{Takahiro Kashiwazaki}
\affiliation{NTT Device Technology Labs, NTT Inc., Atsugi, Kanagawa 243-0198, Japan}

\author{Asuka Inoue}
\affiliation{NTT Device Technology Labs, NTT Inc., Atsugi, Kanagawa 243-0198, Japan}

\author{Takeshi Umeki}
\affiliation{NTT Device Technology Labs, NTT Inc., Atsugi, Kanagawa 243-0198, Japan}

\author{Petr Marek}
\affiliation{Department of Optics, Palacky University, 17. listopadu 1192/12, Olomouc 77146, Czech Republic}

\author{Radim Filip}
\affiliation{Department of Optics, Palacky University, 17. listopadu 1192/12, Olomouc 77146, Czech Republic}

\author{Warit Asavanant}
\affiliation{Department of Applied Physics, School of Engineering, The University of Tokyo, Tokyo 113-8656, Japan}
\affiliation{Optical Quantum Computing Research Team, RIKEN Center for Quantum Computing, Wako, Saitama 351-0198, Japan}
\affiliation{OptQC Corp., 1-21-7 Nishi-Ikebukuro, Toshima, Tokyo, Japan}

\author{Akira Furusawa}
\email{akiraf@ap.t.u-tokyo.ac.jp}
\affiliation{Department of Applied Physics, School of Engineering, The University of Tokyo, Tokyo 113-8656, Japan}
\affiliation{Optical Quantum Computing Research Team, RIKEN Center for Quantum Computing, Wako, Saitama 351-0198, Japan}
\affiliation{OptQC Corp., 1-21-7 Nishi-Ikebukuro, Toshima, Tokyo, Japan}

\date{\today}

\begin{abstract}
Wave--particle duality is a hallmark of quantum mechanics. For bosonic systems, there exists a continuum of intermediate states bridging wave-like Schr\"odinger cat states and particle-like Fock states. Such states have recently been recognized as valuable resources for enhancing fault-tolerant quantum computation (FTQC) with propagating light. Here we experimentally demonstrate tunable generation of these intermediate states by employing generalized photon subtraction (GPS). By detecting up to three photons from squeezed-light sources with a photon-number-resolving detector, we continuously control the balance between wave- and particle-like features. This approach allows us to construct a spectral family of quantum states with high generation rates, optimized according to the required fault-tolerance threshold. Our results establish GPS as a versatile toolbox for tailoring non-Gaussian resources, opening a pathway to efficient Gottesman--Kitaev--Preskill (GKP) qubit generation and addressing a central bottleneck in optical quantum computing.
\end{abstract}

\maketitle

Wave--particle duality is one of the most iconic and paradoxical phenomena in quantum mechanics \cite{Bohr1928,Greenberger1988, Englert1996}, describing the simultaneous manifestation of wave-like and particle-like properties in a single quantum system. In optics, Young's double-slit experiment revealed spatial interference as evidence of wave behavior, while the photoelectric effect established photons as discrete particles \cite{Taylor1909, Tonomura1989, Young1997, Aspden2016}. Today, such duality is recognized as a universal property of microscopic matter and fields \cite{Einstein1905, Millikan1916, Lenard1902}.

Beyond its foundational importance, wave--particle duality in a phase space of optical amplitude can be harnessed as a novel engineering degree of freedom for quantum technologies. In optical quantum states, the particle aspect is exemplified by Fock states (or photon-number states) $\lvert n\rangle$, while the wave aspect is embodied by Schr\"odinger cat states (or superposition of coherent states). Both serve as indispensable resources for quantum sensing \cite{Munro2002, Zurek2001, Birrittella2021, Holland1993, Giovannetti2006}, communication \cite{vanEnk2001, Sangouard2010,vanLoock2008,Gisin2002}, and computation \cite{Cochrane1999,Ralph2003,Guillaud2019,Knill2001, Bartolucci2023, Wang2019}. Yet the Hilbert space of single-mode optical states is not confined to these extremes: a continuous family of intermediate states exists between them with a non-Gaussian interference in phase space. 

\begin{figure*}[t]
    \centering
    \includegraphics[width=0.8\textwidth]{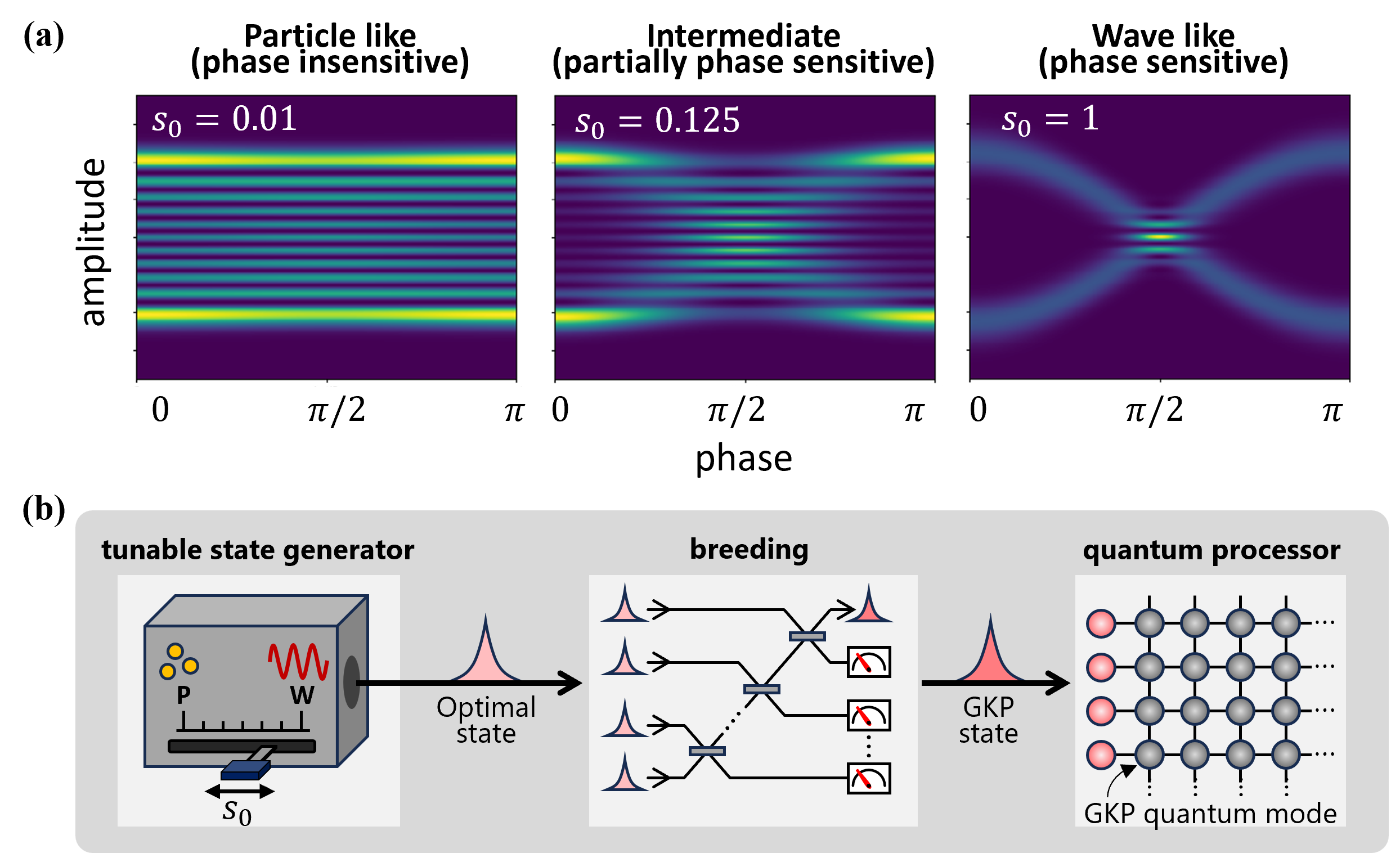}
    \caption{
         Simulation and applications of generalized photon subtraction (GPS). 
         (a) Phase-resolved quadrature marginal distributions, showing the continuous transition from wave-like (phase-sensitive), through intermediate, to particle-like (phase-insensitive) states. 
        (b) Conceptual overview of the role of GPS in optical quantum computing: a tunable state generator based on GPS produces intermediate states that serve as inputs for a breeding protocol, yielding logical (GKP) states for an optical FTQC \cite{takase2024}. }
    \label{fig1}
\end{figure*}

Generalized photon subtraction (GPS) provides direct access to this diverse continuum \cite{Takase2021}. GPS is a state generation method leveraging photon-number-resolving detection. When $n$ photons are detected in one arm of a two-mode Gaussian entangled state, the conditional wavefunction in the other arm is given by
\begin{equation}
\Psi_{n,s_0}(x) \propto (\phi_0(x))^{s_0}\,\phi_n(x),
\end{equation}
where $\phi_n(x)=\frac{1}{\pi^{1/4}\sqrt{2^n n!}}H_n(x)e^{-x^2/2}$ denotes the harmonic-oscillator wavefunction of the Fock state $\lvert n\rangle$ with $H_n(x)$ being the Hermite polynomial, and the parameter $s_0$ governs the balance between wave- and particle-like character (also known as non-Gaussian phase sensitivity \cite{Hanamura2025}). An arbitrary value of $s_0$ ($0 \leq s_0 < \infty$) can be realized by choosing appropriate initial Gaussian entangled states, which, as shown later in this paper, can be readily implemented with experimentally accessible operations.
The resulting state evolution as a function of $s_0$ is shown in Fig.~\ref{fig1}(a). For large $s_0$, the state exhibits a cat-like character, characterized by a bimodal structure in the quadrature distribution and the presence of quantum-interference fringes in phase space, while in the limit $s_0 \to 0$ it converges to a pure Fock state. Intermediate values of $s_0$ therefore correspond to genuinely hybrid states, enabling continuous tuning between wave- and particle-like character. Previously demonstrated non-Gaussian state generation schemes based on photon subtraction are confined to the large-$s_0$ (cat-like) regime \cite{Danka1997, Ourjoumtsev2006, Neergaard-Nielsen2006, Endo2021}. This is because photon-subtraction-based cat-state generation can, in principle, only produce states with $s_0 > 1$. In contrast, GPS provides access to the full $s_0$ range, offering a versatile platform for generating a broad class of non-Gaussian quantum resources.

 This tunability is not only of conceptual significance but also of practical importance. In particular, Gottesman--Kitaev--Preskill (GKP) qubits \cite{Gottesman2001, Fluhmann2019, Campagne-Ibarcq2020, Konno2024, Larsen2025}---leading candidates for fault-tolerant quantum computation (FTQC)---can be generated more efficiently when optimal intermediate states are employed. For GKP state generation, the breeding approach \cite{Vasconcelos2010}, in which multiple resource states are interfered and post-selected, is a promising method. The key metrics for GKP state generation, such as the minimum required detected photon number and the overall success probability, depend strongly on the choice of resource states. Intermediate states $\Psi_{n,s_0}(x)$ ($s_0\sim1$) provide the most favorable performance as resource states, the optimal value of $s_0$ further depending on the desired GKP state quality \cite{Hanamura2025}. An overview of an optical quantum computer using intermediate states is shown in Fig.~\ref{fig1}(b). While breeding protocols \cite{Sychev2017,Konno2024} and optical quantum processors \cite{Aghaee2025,Yokoyama2025} have already been demonstrated, as mentioned, in these experiments, only the cat-like states with $s_0>1$ were used as the input resource states. Due to the constraint on the resource states, breeding-based generation requires excessive photon detections and results in extremely low success probabilities \cite{Fiurášek2025}. Therefore, developing techniques to generate intermediate states with smaller $s_0$ is a natural and essential direction.

In this work, we experimentally demonstrate GPS-based generation of intermediate states by combining squeezed-light sources with an arrayed superconducting-nanostrip photon detector (SNSPD). By conditioning on up to three detected photons, and continuously tuning the parameter $s_0$, we mapped out the full spectrum of wave--particle duality. This establishes a ``tunable state generator'' for FTQC. Our results thus provide a crucial step toward overcoming the bottleneck of resource generation in optical quantum computing and open a pathway to the early realization of FTQC with light.

\begin{figure*}[t]
    \centering
    \includegraphics[width=0.8\textwidth]{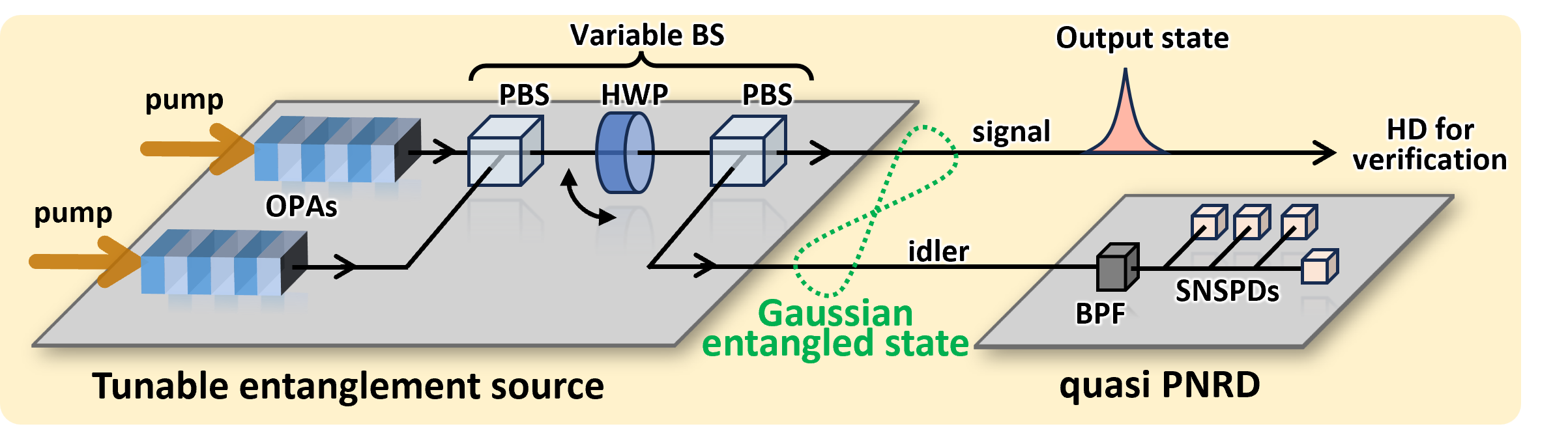}
    \caption{
         Experimental setup. Two broadband squeezed-vacuum sources (OPAs) are interfered with on a variable beam splitter (PBS--HWP--PBS). Conditional detection of up to three photons with SNSPDs, after spectral filtering with the bandwidth of \SI{15}{MHz}, heralds intermediate non-Gaussian states, which are then characterized by homodyne detection (HD).}
    \label{fig2}
\end{figure*}

We now describe the experimental implementation of GPS \cite{Takase2021}. We constructed an optical system capable of continuously controlling wave--particle duality, as shown in Fig.~\ref{fig2}. A continuous-wave laser at \SI{1545.32}{nm} was amplified and frequency-doubled to generate \SI{772.66}{nm} pump light. Two independent squeezed-vacuum modes were produced by injecting this pump into 45-mm-long periodically-poled lithium niobate (PPLN) waveguides, with the parametric gains of 2.11 and 2.05, respectively. These modes were then interfered with on a variable beam splitter (variable BS) composed of two polarization beam splitters (PBSs) and a half-wave plate (HWP). The reflectivity of the variable BS was continuously tunable by rotating the HWP. This part of the setup can be regarded as a tunable entanglement source. In other words, by tuning the reflectivity of the variable beam splitter, we directly control the parameter $s_0$, which governs the balance between wave- and particle-like character.

The two interfered outputs were designated as the ``idler'' and ``signal'' modes. The idler was passed through a spectral filter with a bandwidth of \SI{15}{MHz} and coupled into a single-mode fiber that distributed the light to the arrayed SNSPD. The SNSPD array consisted of four detectors, and their detection efficiencies and dark-count rates were as follows: SNSPD1: \SI{75}{\%}, \SI{40}{cps}; SNSPD2: \SI{67}{\%}, \SI{97}{cps}; SNSPD3: \SI{75}{\%}, \SI{26}{cps}; SNSPD4: \SI{62}{\%}, \SI{32}{cps}. Detection of up to three photons heralded the conditional generation of the corresponding quantum state in the signal mode.

The quadrature amplitudes of the signal mode were verified by balanced homodyne detection (HD) using two high-efficiency photodiodes and a low-noise transimpedance amplifier. The phase of the local-oscillator beam for homodyne detection was switched using an electro-optic phase modulator, and homodyne data were acquired at measurement bases of 0, 30, 60, 90, 120, and 150 degrees, with 20,000 samples recorded for each setting. Phase stabilization of both the interferometer and the homodyne reference was achieved by means of a co-propagating probe beam.

For each measurement phase, we acquired homodyne samples and reconstructed the quantum states using maximum-likelihood estimation \cite{Lvovsky2004}, yielding density matrices, Wigner functions, and marginal distributions.

\begin{figure*}[t]
    \centering
    \includegraphics[width=0.95\textwidth]{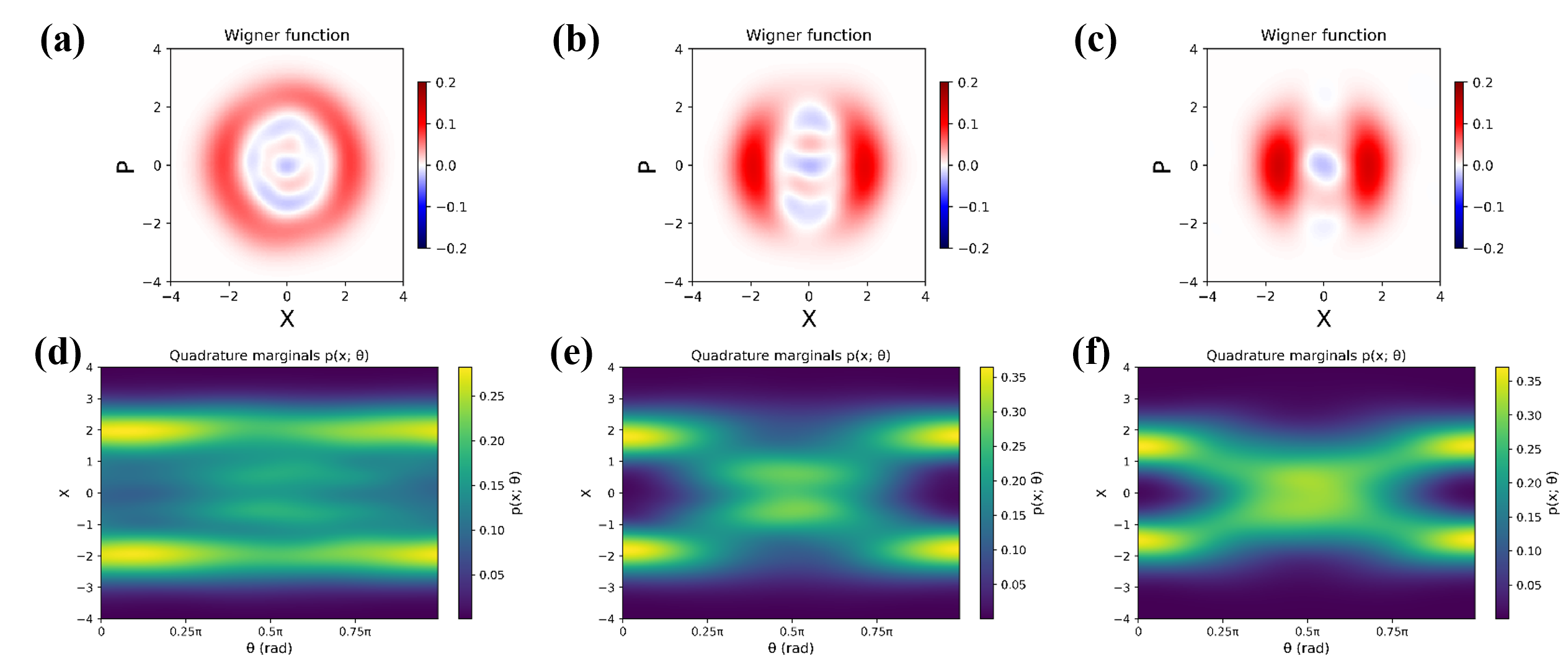}
    \caption{%
        Reconstructed quantum states generated by GPS. 
        Top row (a - c): Wigner functions of representative cases---(a) Fock-like state ($s_0=0.11, R=50\%$), (b) intermediate state ($s_0=0.5, R=40\%$), and (c) cat-like state ($s_0=1.2, R=30\%$). 
        All states exhibit negativity, confirming their strong nonclassicality. 
        Bottom row (d - f): Phase-resolved quadrature marginal distributions $p(x;\theta)$. 
        The cat-like and intermediate states display phase-dependent oscillatory features, while the Fock-like state remains nearly phase-insensitive, illustrating the tunable transition from wave- to particle-like behavior, respectively.
    }
    \label{fig3}
\end{figure*}

We fixed the relative phase of the squeezed-light inputs at the variable beam splitter to $90^\circ$ and performed measurements while tuning only the variable BS reflectivity. This adjustment effectively controlled the parameter $s_0$, which governs the balance between particle-like and wave-like features in the generated states. Representative examples are shown in Fig.~\ref{fig3}. A Fock-like state exhibiting strong particle character ($s_0=0.11$, reflectivity 50\%), a cat-like state dominated by wave interference ($s_0=1.2$, reflectivity 30\%), and an intermediate state with $s_0 = 0.5$ in which both features coexist (reflectivity 40\%). In other words, in the intermediate regime, the quadrature distribution exhibits partial phase sensitivity while simultaneously showing phase-insensitive features, indicating the coexistence of wave-like interference and particle-like discreteness. The corresponding event rates at these reflectivities were \SI{8.6}{cps}, \SI{12.2}{cps}, and \SI{9.3}{cps}, respectively.

The reconstructed Wigner functions reveal several important features, as visible in Fig.~\ref{fig3}(a--c). First, all states exhibit three pronounced negativities without loss correction, as expected by the ideal three-photon measurement, demonstrating their highly non-Gaussian character. Second, the Fock-like state shows circular symmetry in phase space and reduced phase sensitivity, consistent with the expected behavior of a number state. In contrast, both the cat and intermediate states break rotational symmetry and exhibit phase-dependent interference fringes in their Wigner functions. As $s_0$ increases, the fringe contrast increases, indicating the gradual enhancement of wave-like interference. These qualitative differences are also reflected in the marginal quadrature distributions and in the cuts of the Wigner function $W(x,0)$ and $W(0,p)$ in Fig.~\ref{fig3}(a--c). As shown in Fig.~\ref{fig3}(d--f), the particle-like state yields nearly identical histograms for all local-oscillator phases, whereas the intermediate and cat states display increasingly pronounced oscillatory patterns, directly visualizing the transition from particle-dominated to wave-dominated behavior.

These results establish generalized photon subtraction as an experimental platform for continuously controlling wave--particle duality in non-Gaussian optical states. Unlike demonstrations of quantum duality based on spatial or temporal interference \cite{Taylor1909}, the present work realizes duality as an intrinsic, tunable property of a single-mode quantum state, controlled by a well-defined experimental parameter. This capability goes beyond a conceptual illustration of duality and provides a practical tool for engineering non-Gaussian quantum resources.

In the context of FTQC, GKP qubits represent a leading logical encoding for continuous-variable systems. However, their preparation remains a central experimental challenge. Previous demonstrations of cat-state breeding relied on multiple squeezed-light sources and high-order photon detections, resulting in success probabilities far below the threshold required for scalable operation. By contrast, theoretical studies have shown that adaptive breeding schemes based on states generated by generalized photon subtraction can substantially enhance the success probability \cite{takase2024}. From this perspective, the present work provides an experimental realization of a key building block: a continuously tunable non-Gaussian resource-state generator that can supply the intermediate states required for optimized GKP synthesis.

Another important aspect of the GPS-based platform is its simplicity and modularity. Each GPS unit consists only of two squeezed-vacuum sources, a single beam splitter, and a photon-number-resolving detector. This minimal unit cell arises naturally from the fact that wave--particle duality is controlled by a single experimental parameter, and it is well suited to architectures in which circuit complexity and fabrication yield are critical considerations. Moreover, the Gaussian operations required in adaptive breeding protocols---such as conditional squeezing based on the detected photon number---can, in principle, be implemented deterministically using measurement-based squeezing techniques \cite{Filip2005,Miwa2014}. Together, these features position generalized photon subtraction as a practical and flexible platform for supplying non-Gaussian resources in photonic quantum information processing.

In conclusion, we have experimentally demonstrated that generalized photon subtraction enables continuous tuning of wave--particle duality in quantum light. By combining squeezed-light sources with photon-number-resolving detection, we realized a single experimental platform in which the balance between particle-like and wave-like features can be continuously controlled through the parameter $s_0$. The reconstructed Wigner functions exhibit clear negativities for all operating points, confirming strong non-Gaussianity without loss correction. Importantly, the degree of phase sensitivity evolves smoothly as a function of $s_0$, directly visualizing the continuous interpolation between particle-dominated and wave-dominated regimes. These results establish wave--particle duality as a controllable resource in the phase space of optical quantum states.

\begin{acknowledgments}
This work was partly supported by Japan Science and Technology (JST) Agency (Moonshot R\&D, Grant No. JPMJMS2064, and PRESTO, Grant No. JPMJPR2254), the UTokyo Foundation, and donations from Nichia Corporation. W.A. acknowledges the funding from Japan Society for the Promotion of Science (JSPS) KAKENHI (Grant No. 23K13040). M.E., K.T, and W.A. acknowledge supports from the Research Foundation for Opto-Science and Technology. P.M. acknowledges the project 25-17472S of the Czech Science Foundation, the European Union’s HORIZON Research and Innovation Actions under Grant Agreement no. 101080173 (CLUSTEC) and the project CZ.02.01.01\/00\/22\_008\/0004649 (QUEENTEC) of EU and the Czech Ministry of Education, Youth and Sport. R.F. acknowledges the project No. 21-13265X of the Czech Science Foundation.
\end{acknowledgments}

\bibliography{apssamp}

@article{Taylor1909,
   author = {Taylor, G. I.},
   title = {Interference Fringes with Feeble Light},
   journal = {Mathematical Proceedings of the Cambridge Philosophical Society},
   volume = {15},
   pages = {114-115},
   year = {1909},
   type = {Journal Article}
}

@article{Fiurášek2025,
   author = {Fiurášek, Jaromír},
   title = {Maximum heralding probabilities of non-classical state generation from two-mode Gaussian state via photon counting measurements},
   journal = {arXiv},
   pages = {arXiv:2510.01951},
   DOI = {10.48550/arXiv.2510.01951},
   year = {2025},
   type = {Journal Article}
}

@article{Miwa2014,
   author = {Miwa, Y. and Yoshikawa, J. and Iwata, N. and Endo, M. and Marek, P. and Filip, R. and van Loock, P. and Furusawa, A.},
   title = {Exploring a new regime for processing optical qubits: squeezing and unsqueezing single photons},
   journal = {Phys Rev Lett},
   volume = {113},
   number = {1},
   pages = {013601},
   ISSN = {1079-7114 (Electronic)
0031-9007 (Linking)},
   DOI = {10.1103/PhysRevLett.113.013601},
   url = {https://www.ncbi.nlm.nih.gov/pubmed/25032925},
   year = {2014},
   type = {Journal Article}
}

@article{Sychev2017,
   author = {Sychev, Demid V. and Ulanov, Alexander E. and Pushkina, Anastasia A. and Richards, Matthew W. and Fedorov, Ilya A. and Lvovsky, Alexander I.},
   title = {Enlargement of optical Schrödinger's cat states},
   journal = {Nature Photonics},
   volume = {11},
   number = {6},
   pages = {379-382},
   ISSN = {1749-4885
1749-4893},
   DOI = {10.1038/nphoton.2017.57},
   year = {2017},
   type = {Journal Article}
}

@article{Yokoyama2025,
   author = {Yokoyama, Shota and Sakaguchi, Atsushi and Asavanant, Warit and Takase, Kan and Chen, Yi-Ru and Nagayoshi, Hironari and Yoshikawa, Jun-ichi and Kashiwazaki, Takahiro and Inoue, Asuka and Umeki, Takeshi and Hashimoto, Toshikazu and Hiraoka, Takuji and Furusawa, Akira and Yonezawa, Hidehiro},
   title = {Full-stack Analog Optical Quantum Computer with A Hundred Inputs},
   journal = {arXiv},
   pages = {arXiv:2506.16147 [quant-ph]},
   DOI = {https://doi.org/10.48550/arXiv.2506.16147},
   year = {2025},
   type = {Journal Article}
}

@article{Aghaee2025,
   author = {Aghaee Rad, H. and Ainsworth, T. and Alexander, R. N. and Altieri, B. and Askarani, M. F. and Baby, R. and Banchi, L. and Baragiola, B. Q. and Bourassa, J. E. and Chadwick, R. S. and Charania, I. and Chen, H. and Collins, M. J. and Contu, P. and D'Arcy, N. and Dauphinais, G. and De Prins, R. and Deschenes, D. and Di Luch, I. and Duque, S. and Edke, P. and Fayer, S. E. and Ferracin, S. and Ferretti, H. and Gefaell, J. and Glancy, S. and Gonzalez-Arciniegas, C. and Grainge, T. and Han, Z. and Hastrup, J. and Helt, L. G. and Hillmann, T. and Hundal, J. and Izumi, S. and Jaeken, T. and Jonas, M. and Kocsis, S. and Krasnokutska, I. and Larsen, M. V. and Laskowski, P. and Laudenbach, F. and Lavoie, J. and Li, M. and Lomonte, E. and Lopetegui, C. E. and Luey, B. and Lund, A. P. and Ma, C. and Madsen, L. S. and Mahler, D. H. and Mantilla Calderon, L. and Menotti, M. and Miatto, F. M. and Morrison, B. and Nadkarni, P. J. and Nakamura, T. and Neuhaus, L. and Niu, Z. and Noro, R. and Papirov, K. and Pesah, A. and Phillips, D. S. and Plick, W. N. and Rogalsky, T. and Rortais, F. and Sabines-Chesterking, J. and Safavi-Bayat, S. and Sazhaev, E. and Seymour, M. and Rezaei Shad, K. and Silverman, M. and Srinivasan, S. A. and Stephan, M. and Tang, Q. Y. and Tasker, J. F. and Teo, Y. S. and Then, R. B. and Tremblay, J. E. and Tzitrin, I. and Vaidya, V. D. and Vasmer, M. and Vernon, Z. and Villalobos, Lfssm and Walshe, B. W. and Weil, R. and Xin, X. and Yan, X. and Yao, Y. and Zamani Abnili, M. and Zhang, Y.},
   title = {Scaling and networking a modular photonic quantum computer},
   journal = {Nature},
   ISSN = {1476-4687 (Electronic)
0028-0836 (Linking)},
   DOI = {10.1038/s41586-024-08406-9},
   url = {https://www.ncbi.nlm.nih.gov/pubmed/39843755},
   year = {2025},
   type = {Journal Article}
}

@article{Hanamura2025,
   author = {Hanamura, F. and Takase, K. and Nagayoshi, H. and Ide, R. and Asavanant, W. and Fukui, K. and Marek, P. and Filip, R. and Furusawa, A.},
   title = {Beyond Stellar Rank: Control Parameters for Scalable Optical Non-Gaussian State Generation},
   journal = {arXiv},
   year = {2025},
   pages = {arXiv:2509.06255v1},
   DOI = {10.48550/arXiv.2509.06255},
   type = {Journal Article}
}

@article{Endo2021,
   author = {Endo, M. and Sonoyama, T. and Matsuyama, M. and Okamoto, F. and Miki, S. and Yabuno, M. and China, F. and Terai, H. and Furusawa, A.},
   title = {Quantum detector tomography of a superconducting nanostrip photon-number-resolving detector},
   journal = {Opt Express},
   volume = {29},
   number = {8},
   pages = {11728-11738},
   ISSN = {1094-4087 (Electronic)
1094-4087 (Linking)},
   DOI = {10.1364/OE.423142},
   url = {https://www.ncbi.nlm.nih.gov/pubmed/33984948},
   year = {2021},
   type = {Journal Article}
}

@article{Aspden2016,
   author = {Aspden, Reuben S. and Padgett, Miles J. and Spalding, Gabriel C.},
   title = {Video recording true single-photon double-slit interference},
   journal = {American Journal of Physics},
   volume = {84},
   number = {9},
   pages = {671-677},
   ISSN = {0002-9505
1943-2909},
   DOI = {10.1119/1.4955173},
   year = {2016},
   type = {Journal Article}
}

@article{Bartolucci2023,
   author = {Bartolucci, S. and Birchall, P. and Bombin, H. and Cable, H. and Dawson, C. and Gimeno-Segovia, M. and Johnston, E. and Kieling, K. and Nickerson, N. and Pant, M. and Pastawski, F. and Rudolph, T. and Sparrow, C.},
   title = {Fusion-based quantum computation},
   journal = {Nat Commun},
   volume = {14},
   number = {1},
   pages = {912},
   ISSN = {2041-1723 (Electronic)
2041-1723 (Linking)},
   DOI = {10.1038/s41467-023-36493-1},
   url = {https://www.ncbi.nlm.nih.gov/pubmed/36805650},
   year = {2023},
   type = {Journal Article}
}

@article{Birrittella2021,
   author = {Birrittella, Richard J. and Alsing, Paul M. and Gerry, Christopher C.},
   title = {The parity operator: Applications in quantum metrology},
   journal = {AVS Quantum Science},
   volume = {3},
   number = {1},
   ISSN = {2639-0213},
   DOI = {10.1116/5.0026148},
   year = {2021},
   type = {Journal Article}
}

@article{Bohr1928,
   author = {Bohr, N.},
   title = {The Quantum Postulate and the Recent Development of Atomic Theory1},
   journal = {Nature},
   volume = {121},
   number = {3050},
   pages = {580-590},
   ISSN = {1476-4687},
   DOI = {10.1038/121580a0},
   url = {https://doi.org/10.1038/121580a0},
   year = {1928},
   type = {Journal Article}
}

@article{Campagne-Ibarcq2020,
   author = {Campagne-Ibarcq, P. and Eickbusch, A. and Touzard, S. and Zalys-Geller, E. and Frattini, N. E. and Sivak, V. V. and Reinhold, P. and Puri, S. and Shankar, S. and Schoelkopf, R. J. and Frunzio, L. and Mirrahimi, M. and Devoret, M. H.},
   title = {Quantum error correction of a qubit encoded in grid states of an oscillator},
   journal = {Nature},
   volume = {584},
   number = {7821},
   pages = {368-372},
   ISSN = {1476-4687 (Electronic)
0028-0836 (Linking)},
   DOI = {10.1038/s41586-020-2603-3},
   url = {https://www.ncbi.nlm.nih.gov/pubmed/32814889},
   year = {2020},
   type = {Journal Article}
}

@article{Cochrane1999,
   author = {Cochrane, P. T. and Milburn, G. J. and Munro, W. J.},
   title = {Macroscopically distinct quantum-superposition states as a bosonic code for amplitude damping},
   journal = {Physical Review A},
   volume = {59},
   number = {4},
   pages = {2631-2634},
   ISSN = {1050-2947
1094-1622},
   DOI = {10.1103/PhysRevA.59.2631},
   year = {1999},
   type = {Journal Article}
}

@article{Danka1997,
   author = {Dakna, M. and Anhut, T. and Opatrný, T. and Knöll, L. and Welsch, D. G.},
   title = {Generating Schrödinger-cat-like states by means of conditional measurements on a beam splitter},
   journal = {Physical Review A},
   volume = {55},
   number = {4},
   pages = {3184-3194},
   ISSN = {1050-2947
1094-1622},
   DOI = {10.1103/PhysRevA.55.3184},
   year = {1997},
   type = {Journal Article}
}

@article{Einstein1905,
   author = {Einstein, A.},
   title = {Über einen die Erzeugung und Verwandlung des Lichtes betreffenden heuristischen Gesichtspunkt},
   journal = {Annalen der Physik},
   volume = {322},
   number = {6},
   pages = {132-148},
   ISSN = {0003-3804
1521-3889},
   DOI = {10.1002/andp.19053220607},
   year = {1905},
   type = {Journal Article}
}

@article{Englert1996,
   author = {Englert, B. G.},
   title = {Fringe Visibility and Which-Way Information: An Inequality},
   journal = {Phys Rev Lett},
   volume = {77},
   number = {11},
   pages = {2154-2157},
   ISSN = {1079-7114 (Electronic)
0031-9007 (Linking)},
   DOI = {10.1103/PhysRevLett.77.2154},
   url = {https://www.ncbi.nlm.nih.gov/pubmed/10061872},
   year = {1996},
   type = {Journal Article}
}

@article{Fluhmann2019,
   author = {Fluhmann, C. and Nguyen, T. L. and Marinelli, M. and Negnevitsky, V. and Mehta, K. and Home, J. P.},
   title = {Encoding a qubit in a trapped-ion mechanical oscillator},
   journal = {Nature},
   volume = {566},
   number = {7745},
   pages = {513-517},
   ISSN = {1476-4687 (Electronic)
0028-0836 (Linking)},
   DOI = {10.1038/s41586-019-0960-6},
   url = {https://www.ncbi.nlm.nih.gov/pubmed/30814715},
   year = {2019},
   type = {Journal Article}
}

@article{Giovannetti2006,
   author = {Giovannetti, V. and Lloyd, S. and Maccone, L.},
   title = {Quantum metrology},
   journal = {Phys Rev Lett},
   volume = {96},
   number = {1},
   pages = {010401},
   ISSN = {0031-9007 (Print)
0031-9007 (Linking)},
   DOI = {10.1103/PhysRevLett.96.010401},
   url = {https://www.ncbi.nlm.nih.gov/pubmed/16486424},
   year = {2006},
   type = {Journal Article}
}

@article{Gisin2002,
   author = {Gisin, Nicolas and Ribordy, Grégoire and Tittel, Wolfgang and Zbinden, Hugo},
   title = {Quantum cryptography},
   journal = {Reviews of Modern Physics},
   volume = {74},
   number = {1},
   pages = {145-195},
   ISSN = {0034-6861
1539-0756},
   DOI = {10.1103/RevModPhys.74.145},
   year = {2002},
   type = {Journal Article}
}

@article{Gottesman2001,
   author = {Gottesman, Daniel and Kitaev, Alexei and Preskill, John},
   title = {Encoding a qubit in an oscillator},
   journal = {Physical Review A},
   volume = {64},
   number = {1},
   pages = {012310},
   ISSN = {1050-2947
1094-1622},
   DOI = {10.1103/PhysRevA.64.012310},
   year = {2001},
   type = {Journal Article}
}

@article{Greenberger1988,
   author = {Greenberger, Daniel M. and Yasin, Allaine},
   title = {Simultaneous wave and particle knowledge in a neutron interferometer},
   journal = {Physics Letters A},
   volume = {128},
   number = {8},
   pages = {391-394},
   ISSN = {03759601},
   DOI = {10.1016/0375-9601(88)90114-4},
   year = {1988},
   type = {Journal Article}
}

@article{Guillaud2019,
   author = {Guillaud, Jérémie and Mirrahimi, Mazyar},
   title = {Repetition Cat Qubits for Fault-Tolerant Quantum Computation},
   journal = {Physical Review X},
   volume = {9},
   number = {4},
   ISSN = {2160-3308},
   DOI = {10.1103/PhysRevX.9.041053},
   year = {2019},
   type = {Journal Article}
}

@article{Holland1993,
   author = {Holland, M. J. and Burnett, K.},
   title = {Interferometric detection of optical phase shifts at the Heisenberg limit},
   journal = {Phys Rev Lett},
   volume = {71},
   number = {9},
   pages = {1355-1358},
   ISSN = {1079-7114 (Electronic)
0031-9007 (Linking)},
   DOI = {10.1103/PhysRevLett.71.1355},
   url = {https://www.ncbi.nlm.nih.gov/pubmed/10055519},
   year = {1993},
   type = {Journal Article}
}

@article{Knill2001,
   author = {Knill, E. and Laflamme, R. and Milburn, G. J.},
   title = {A scheme for efficient quantum computation with linear optics},
   journal = {Nature},
   volume = {409},
   number = {6816},
   pages = {46-52},
   ISSN = {0028-0836 (Print)
0028-0836 (Linking)},
   DOI = {10.1038/35051009},
   url = {https://www.ncbi.nlm.nih.gov/pubmed/11343107},
   year = {2001},
   type = {Journal Article}
}

@article{Konno2024,
   author = {Konno, S. and Asavanant, W. and Hanamura, F. and Nagayoshi, H. and Fukui, K. and Sakaguchi, A. and Ide, R. and China, F. and Yabuno, M. and Miki, S. and Terai, H. and Takase, K. and Endo, M. and Marek, P. and Filip, R. and van Loock, P. and Furusawa, A.},
   title = {Logical states for fault-tolerant quantum computation with propagating light},
   journal = {Science},
   volume = {383},
   number = {6680},
   pages = {289-293},
   ISSN = {1095-9203 (Electronic)
0036-8075 (Linking)},
   DOI = {10.1126/science.adk7560},
   url = {https://www.ncbi.nlm.nih.gov/pubmed/38236963},
   year = {2024},
   type = {Journal Article}
}

@article{Larsen2025,
   author = {Larsen, M. V. and Bourassa, J. E. and Kocsis, S. and Tasker, J. F. and Chadwick, R. S. and Gonzalez-Arciniegas, C. and Hastrup, J. and Lopetegui-Gonzalez, C. E. and Miatto, F. M. and Motamedi, A. and Noro, R. and Roeland, G. and Baby, R. and Chen, H. and Contu, P. and Di Luch, I. and Drago, C. and Giesbrecht, M. and Grainge, T. and Krasnokutska, I. and Menotti, M. and Morrison, B. and Puviraj, C. and Rezaei Shad, K. and Hussain, B. and McMahon, J. and Ortmann, J. E. and Collins, M. J. and Ma, C. and Phillips, D. S. and Seymour, M. and Tang, Q. Y. and Yang, B. and Vernon, Z. and Alexander, R. N. and Mahler, D. H.},
   title = {Integrated photonic source of Gottesman-Kitaev-Preskill qubits},
   journal = {Nature},
   volume = {642},
   number = {8068},
   pages = {587-591},
   ISSN = {1476-4687 (Electronic)
0028-0836 (Print)
0028-0836 (Linking)},
   DOI = {10.1038/s41586-025-09044-5},
   url = {https://www.ncbi.nlm.nih.gov/pubmed/40468070},
   year = {2025},
   type = {Journal Article}
}

@article{Lenard1902,
   author = {Lenard, P.},
   title = {Ueber die lichtelektrische Wirkung},
   journal = {Annalen der Physik},
   volume = {313},
   number = {5},
   pages = {149-198},
   ISSN = {0003-3804
1521-3889},
   DOI = {10.1002/andp.19023130510},
   year = {1902},
   type = {Journal Article}
}

@article{Millikan1916,
   author = {Millikan, R. A.},
   title = {A Direct Photoelectric Determination of Planck's "h"},
   journal = {Physical Review},
   volume = {7},
   number = {3},
   pages = {355-388},
   ISSN = {0031-899X},
   DOI = {10.1103/PhysRev.7.355},
   year = {1916},
   type = {Journal Article}
}

@article{Munro2002,
   author = {Munro, W. J. and Nemoto, K. and Milburn, G. J. and Braunstein, S. L.},
   title = {Weak-force detection with superposed coherent states},
   journal = {Physical Review A},
   volume = {66},
   number = {2},
   ISSN = {1050-2947
1094-1622},
   DOI = {10.1103/PhysRevA.66.023819},
   year = {2002},
   type = {Journal Article}
}

@article{Neergaard-Nielsen2006,
   author = {Neergaard-Nielsen, J. S. and Nielsen, B. M. and Hettich, C. and Molmer, K. and Polzik, E. S.},
   title = {Generation of a superposition of odd photon number states for quantum information networks},
   journal = {Phys Rev Lett},
   volume = {97},
   number = {8},
   pages = {083604},
   ISSN = {0031-9007 (Print)
0031-9007 (Linking)},
   DOI = {10.1103/PhysRevLett.97.083604},
   url = {https://www.ncbi.nlm.nih.gov/pubmed/17026305},
   year = {2006},
   type = {Journal Article}
}

@article{Ourjoumtsev2006,
   author = {Ourjoumtsev, A. and Tualle-Brouri, R. and Laurat, J. and Grangier, P.},
   title = {Generating optical Schrodinger kittens for quantum information processing},
   journal = {Science},
   volume = {312},
   number = {5770},
   pages = {83-6},
   ISSN = {1095-9203 (Electronic)
0036-8075 (Linking)},
   DOI = {10.1126/science.1122858},
   url = {https://www.ncbi.nlm.nih.gov/pubmed/16527930},
   year = {2006},
   type = {Journal Article}
}

@article{Ralph2003,
   author = {Ralph, T. C. and Gilchrist, A. and Milburn, G. J. and Munro, W. J. and Glancy, S.},
   title = {Quantum computation with optical coherent states},
   journal = {Physical Review A},
   volume = {68},
   number = {4},
   ISSN = {1050-2947
1094-1622},
   DOI = {10.1103/PhysRevA.68.042319},
   year = {2003},
   type = {Journal Article}
}

@article{Sangouard2010,
   author = {Sangouard, Nicolas and Simon, Christoph and Gisin, Nicolas and Laurat, Julien and Tualle-Brouri, Rosa and Grangier, Philippe},
   title = {Quantum repeaters with entangled coherent states},
   journal = {Journal of the Optical Society of America B},
   volume = {27},
   number = {6},
   ISSN = {0740-3224
1520-8540},
   DOI = {10.1364/josab.27.00a137},
   year = {2010},
   type = {Journal Article}
}

@article{takase2024,
   author = {Takase, Kan and Hanamura, Fumiya and Nagayoshi, Hironari and Bourassa, J. Eli and Alexander, Rafael N. and Kawasaki, Akito and Asavanant, Warit and Endo, Mamoru and Furusawa, Akira},
   title = {Generation of flying logical qubits using generalized photon subtraction with adaptive Gaussian operations},
   journal = {Physical Review A},
   volume = {110},
   number = {1},
   ISSN = {2469-9926
2469-9934},
   DOI = {10.1103/PhysRevA.110.012436},
   year = {2024},
   type = {Journal Article}
}

@article{Takase2021,
   author = {Takase, Kan and Yoshikawa, J and Asavanant, Warit and Endo, Mamoru and Furusawa, Akira},
   title = {Generation of optical Schrödinger cat states by generalized photon subtraction},
   journal = {Physical Review A},
   volume = {103},
   number = {1},
   pages = {013710},
   ISSN = {2469-9926
2469-9934},
   DOI = {10.1103/PhysRevA.103.013710},
   year = {2021},
   type = {Journal Article}
}

@article{Filip2005,
   author = {Filip, Radim and Marek, Petr and Andersen, Ulrik L.},
   title = {Measurement-induced continuous-variable quantum interactions},
   journal = {Physical Review A},
   volume = {71},
   number = {4},
   ISSN = {1050-2947
1094-1622},
   DOI = {10.1103/PhysRevA.71.042308},
   year = {2005},
   type = {Journal Article}
}

@article{Lvovsky2004,
   author = {Lvovsky, A. I.},
   title = {Iterative maximum-likelihood reconstruction in quantum homodyne tomography},
   journal = {Journal of Optics B: Quantum and Semiclassical Optics},
   volume = {6},
   number = {6},
   pages = {S556-S559},
   ISSN = {1464-4266
1741-3575},
   DOI = {10.1088/1464-4266/6/6/014},
   year = {2004},
   type = {Journal Article}
}

@article{Tonomura1989,
   author = {Tonomura, A. and Endo, J. and Matsuda, T. and Kawasaki, T. and Ezawa, H.},
   title = {Demonstration of single-electron buildup of an interference pattern},
   journal = {American Journal of Physics},
   volume = {57},
   number = {2},
   pages = {117-120},
   ISSN = {0002-9505
1943-2909},
   DOI = {10.1119/1.16104},
   year = {1989},
   type = {Journal Article}
}

@article{vanEnk2001,
   author = {van Enk, S. and Hirota, O.},
   title = {Entangled coherent states: Teleportation and decoherence},
   journal = {Physical Review A},
   volume = {64},
   number = {2},
   ISSN = {1050-2947
1094-1622},
   DOI = {10.1103/PhysRevA.64.022313},
   year = {2001},
   type = {Journal Article}
}

@article{vanLoock2008,
   author = {van Loock, Peter and Lütkenhaus, Norbert and Munro, W. J. and Nemoto, Kae},
   title = {Quantum repeaters using coherent-state communication},
   journal = {Physical Review A},
   volume = {78},
   number = {6},
   ISSN = {1050-2947
1094-1622},
   DOI = {10.1103/PhysRevA.78.062319},
   year = {2008},
   type = {Journal Article}
}

@article{Vasconcelos2010,
   author = {Vasconcelos, H. M. and Sanz, L. and Glancy, S.},
   title = {All-optical generation of states for "Encoding a qubit in an oscillator"},
   journal = {Opt Lett},
   volume = {35},
   number = {19},
   pages = {3261-3},
   ISSN = {1539-4794 (Electronic)
0146-9592 (Linking)},
   DOI = {10.1364/OL.35.003261},
   url = {https://www.ncbi.nlm.nih.gov/pubmed/20890353},
   year = {2010},
   type = {Journal Article}
}

@article{Wang2019,
   author = {Wang, H. and Qin, J. and Ding, X. and Chen, M. C. and Chen, S. and You, X. and He, Y. M. and Jiang, X. and You, L. and Wang, Z. and Schneider, C. and Renema, J. J. and Hofling, S. and Lu, C. Y. and Pan, J. W.},
   title = {Boson Sampling with 20 Input Photons and a 60-Mode Interferometer in a 10;14-Dimensional Hilbert Space},
   journal = {Phys Rev Lett},
   volume = {123},
   number = {25},
   pages = {250503},
   ISSN = {1079-7114 (Electronic)
0031-9007 (Linking)},
   DOI = {10.1103/PhysRevLett.123.250503},
   url = {https://www.ncbi.nlm.nih.gov/pubmed/31922765},
   year = {2019},
   type = {Journal Article}
}

\end{document}